\documentclass{aa}
\usepackage{graphicx}

\begin{document}
\newcommand{\MIC}{\mbox{$\mu$m}}

\title{
Transmission curves and effective refraction indices
of MKO near infrared consortium filters at cryogenic
temperatures
}

\titlerunning{
Characterization of MKO NIR consortium filters
}
\authorrunning{F.~Ghinassi, J.~Licandro, E.~Oliva, et al.}

\author
{
F.~Ghinassi\inst{1},
J.~Licandro\inst{1},
E.~Oliva\inst{2,1},
C.~Baffa\inst{2},
A.~Checcucci\inst{2},
G.~Comoretto\inst{2},
S.~Gennari\inst{2}, and
G.~Marcucci\inst{3}
}
          
\institute{ 
Centro Galileo Galilei \& Telescopio Nazionale Galileo, P.O. Box 565
E-38700 S. Cruz de La Palma, Spain
\and
Osservatorio Astrofisico di Arcetri, Largo E. Fermi 5, I-50125 Firenze, Italy
\and
Universit\`a degli studi di Firenze, dipartimento di Astronomia,
 Largo E. Fermi 3, I-50125 Firenze, Italy
}

\offprints{E. Oliva, e-mail oliva@tng.iac.es }

\date{ Received 29 January 2002; accepted 19 February 2002 }

\abstract {
We report transmission measurements at cryogenic temperatures
for 4  
broad-band filters of the Mauna Kea Observatories (MKO) near-infrared 
filter set and 5 narrow-band filters.
The spectral scans were collected using the multi-mode IR instrument of 
the TNG telescope (NICS) in which these filters are permanently mounted 
and commonly used for astronomical observations. 
We determined the transmission curves at a temperature of 78~K and 
found no significant red--leak up to 2.6 \MIC, the data are available in
electronic form on the TNG web page.
We also estimated the variation of the wavelength response with the incidence 
angle and found it compatible with an
effective refractive index of $\simeq$2 (see Eq.~\ref{eq1}). 
}

\maketitle


\section{Introduction} 
\label{introduction}

Near infrared imaging observations are usually made using a set of
``standard'' filters, the most popular being J, H, K which 
match the 3 regions of clear atmospheric transmission at near infrared
wavelengths, i.e. between 1.1 and 2.5 \MIC. Although these bands
were  first defined long ago (see Johnson \cite{johnson1},
\cite{johnson2}, \cite{johnson3}), no standard recipe exists for 
defining and manufacturing these filters. 
Consequently, different 
filters were used in different observatories and, in several cases,
the filter transmission curve at the operating temperature of IR
instruments ($\simeq$77~K) is not available (see e.g. Moro \& Munari 
\cite{munari}).

\begin{figure*}
\centerline{\resizebox{\hsize}{!}
 {\rotatebox{-90}{\includegraphics{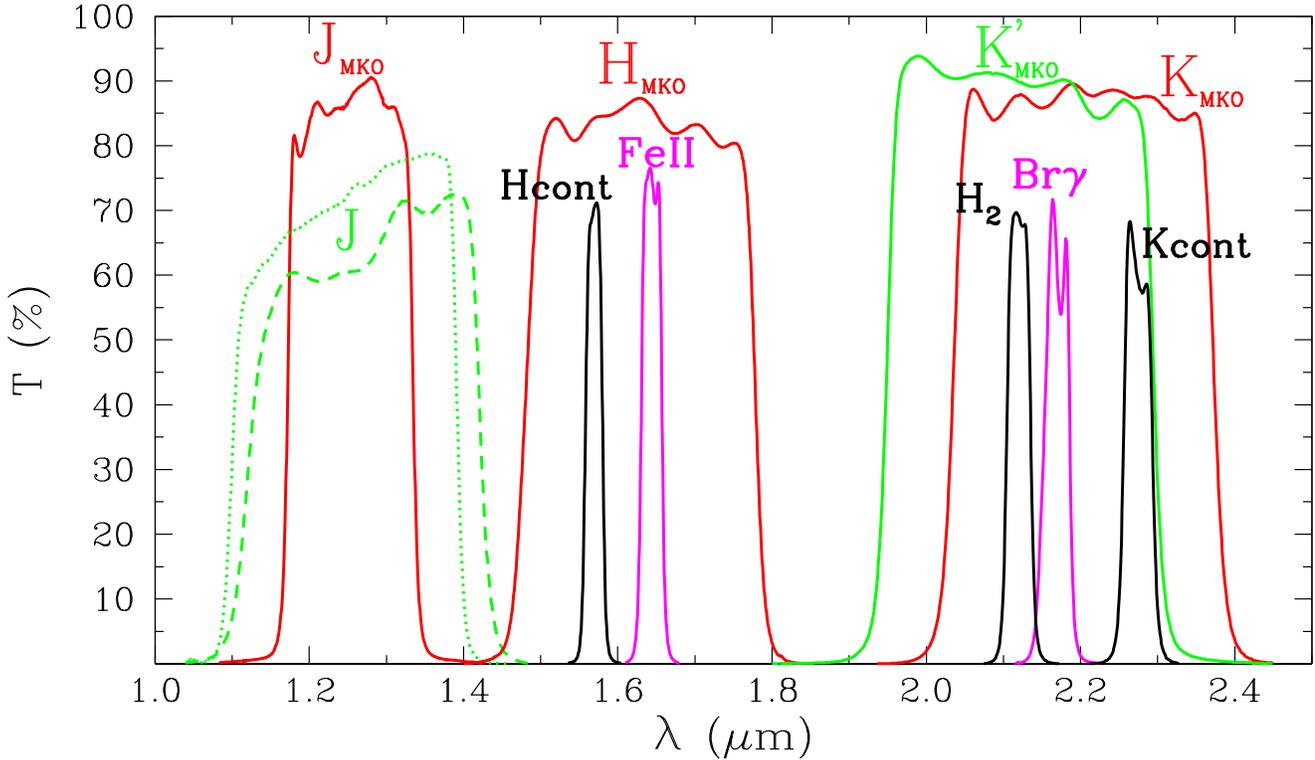}}}}
\caption{
Overview of the transmission curves of the filters of
Mauna Kea Observatories (MKO)
near infrared filter consortium. The data were taken 
at a temperature of 78~K and with the broad and narrow band filters 
illuminated 
at an incidence angle of 5$^o$ and 3$^o$, respectively.
For comparison, we also include the curve relative to the ``standard J''
filter of NICS measured in the same conditions (dashed line) as well as
that of the ``standard J'' filter used in the SOFI and ISAAC
instruments of ESO (dotted line). 
The data for the ESO filter were downloaded from
the ISAAC web page.
}
\label{fig1}
\end{figure*}

Contrary to the broad-band filters at optical wavelengths, several of 
which can be manufactured using combinations of
commercial coloured glasses, those for IR observations can only be produced by
multi-layers coating.
Although some  
firms have produced standard astronomical IR filters for many years, 
the stability of their characteristics
is not always satisfactory. 
An instructive example is comparing the transmission
curves of the
standard J filters manufactured by the same company for ESO and
for the TNG (dotted and dashed curves in Fig.~\ref{fig1}, respectively).

Clarifying this situation has been 
one of the main aims of the MKO
near-infrared filter consortium
(Simons \& Tokunaga \cite{gemini1}, Tokunaga et al. \cite{gemini2}). 
Thanks also to the organizational effort of A. Tokunaga,
a large number of observatories throughout the world could recently
purchase a uniform set of carefully defined, high quality filters at an 
affordable price.

Since direct measurements of the transmission of these filters at cryogenic
temperatures are scarce or, for some of them, lacking altogether, we report
here accurate spectral scans at 78~K for the set of filters purchased for the
multi-mode IR instrument of the Italian 3.5m telescope (TNG). 
The broad-band filters were purchased as part of the first 
MKO filter consortium in 1998 that used OCLI as 
the filter vendor.  The narrow-band filters were purchased as 
part of a separate filter consortium organized in 2000.
We plan to repeat these measurements 
roughly every year to monitor the aging of the coating layers (if any),
the results will be made available in the TNG web page 
({\it www.tng.iac.es/IRfilters.html}) 
where one can also found the data presented
here in electronic form.

\section{Measurements}
\label{measurements}

The data were collected in May 2001 using NICS, the near infrared 
camera and spectrometer which is permanently mounted on the Italian TNG
telescope in La~Palma.
This instrument is a FOSC--type cryogenic focal reducer equipped 
with two interchangeable cameras feeding a Rockwell Hawaii 1024$^2$ array. 
The camera used for the spectroscopic observations has a focal ratio of
F/4.3 and yields a sky projected scale of 0.25"/pixel.
The spectroscopic modes are achieved by means of an Amici prism
and a series of glass--resin
grisms which can be inserted in the $\oslash$ 22 mm collimated beam
(see e.g. Baffa et al. \cite{baffa01}).

All the grisms used for these measurements have their dedicated order sorter
rigidly mounted inside the grism holder. Consequently, normal long-slit
spectroscopic data are collected with the filter wheel in the ``open''
position. Therefore, by inserting a filter in the filter wheel, one can measure
its transmission at the wavelengths covered by the disperser.

The measurements were performed as follows. 
A cold entrance slit of 0.095 mm,
with a projected size of 2 pixels onto the array, was illuminated by an
halogen lamp through a diffuser. After the collimator, the light passed 
through a Lyot stop, the filter and grism wheels before reaching the 
camera and the detector.
A first series of spectra of the halogen lamp were taken by inserting one
of the dispersers available in the grism wheel followed, immediately after,
by a series of spectra with a filter inserted in the filter wheel and, again,
by measurements of the halogen lamp without any filter. Integration times
were identical for all frames and long enough to obtain between 5000 and 30000
$e^-$ per pixels at all the wavelengths of interest, i.e. a good level of
illumination remaining well below the limits of non-linearity.
Each of these cycles lasted a few minutes during which the input flux from
the lamp was found to be stable within better than 1\%. 

The dispersers used were the grisms IJ (0.89--1.45 \MIC, 5.5 \AA/pix),
JH (1.15--1.75 \MIC, 6.6 \AA/pix), JK' (1.15--2.20 \MIC, 11.6 \AA/pix) and
HK (1.40--2.50 \MIC, 11.2 \AA/pix) which allowed determining the filters
transmission with quite fine spectral details. Measurements with
a factor of about 10 lower 
dispersion, i.e. at $\lambda/\Delta\lambda\!\simeq\!100$, were also 
taken with the Amici prism (0.8--2.6 \MIC), these data were particularly 
useful to estimate the out-of-band blocking factor which,
in the red part of the spectrum, and up to array cutoff wavelength of
2.6 \MIC, could be checked to a level of about 0.01\%

The transmission of a given filter was simply determined by dividing frames
taken with and without the filter in the collimated beam. 
Given the very high s/n ratio of the spectra, the internal errors of the
measurements are negligibly small. The actual accuracy of the curves is 
practically limited by systematic effects due, for example, to
the fact that the filters are 
tilted by 3--5 degrees relative to the optical axis and, when they are inserted
in the beam, shift
the pupil image by $\simeq$0.2 mm, i.e. $\simeq$1\% of the pupil diameter.
Therefore, slightly different parts of the dispersers were illuminated 
in the measurements with and without the filter. Nevertheless, each filter
transmission could be independently determined using at least two
different dispersers 
and, consequently, we could estimate that the systematic errors are
$<$4\% in the absolute values and $\le$1\% in the shape of the curves.

Wavelength calibration was performed using exposures of Ar and Xe
lamps which were taken before/after every change of disperser, the
dispersion was always found to be stable within $<$0.1 pixels, i.e.
a factor of $>$20  better than the 
$\lambda/\Delta\lambda\!\simeq\!1000$ resolving power of the spectra.

The variation of the transmission curves with incidence angle was determined
by extracting spectra at different distances for the array center.
The data for narrow band filters spanned a range of
incidence angle from 3$^o$ to 6.5$^o$ while those for broad band filters
extended between 5$^o$ and 7.6$^o$.

\begin{table}
\caption{Summary of filters characteristics$^{(1)}$}

\newcommand{\SKIP}{\noalign{\vskip2pt}}
\begin{flushleft}
\begin{tabular}{lccc}
\hline\hline
\SKIP
Name   & \ \ \ cuton$^{(2)}$ & \ \ \ cutoff$^{(2)}$ 
  & \ \ \  $<\!t\!>^{(3)}$ \\
   &  ($\mu$m) & ($\mu$m) & (\%)  \\
\SKIP
\hline
\SKIP
J$_{\rm MKO}$ & 1.172 & 1.335 & 84  \\
H$_{\rm MKO}$  & 1.481 & 1.779 & 81  \\
K'$_{\rm MKO}$ & 1.949 & 2.295 & 88  \\
K$_{\rm MKO}$  & 2.035 & 2.374 & 85  \\
 & & & \\
Hcont      & 1.5580 & 1.5807 & 67  \\
FeII       & 1.6304 & 1.6576 & 72  \\
H2         & 2.1061 & 2.1376 & 66  \\
Br$\gamma^{(4)}$\ \ \ \ \ \ \ \ \ \ \  & 2.1515 & 2.1868 & 62  \\
Kcont$^{(4)}$    & 2.2551 & 2.2941 & 60  \\

\SKIP
\hline\hline
\end{tabular}
\begin{enumerate}
\item[$^{(1)}$] Measured at $T$=78~K and at an incidence angle of 5$^o$ 
(broad band filters) and 3$^o$ (narrow band filters).
\item[$^{(2)}$] Half-power points
\item[$^{(3)}$] Average transmission between 75\% points
\item[$^{(4)}$] The transmission curve is quite irregular, see Fig.~\ref{fig2}
\end{enumerate} 
\end{flushleft}
\label{tab1}
\end{table}

\section{Results}
\label{results}
\begin{figure}
\centerline{\resizebox{\hsize}{!}
 {\rotatebox{-90}{\includegraphics{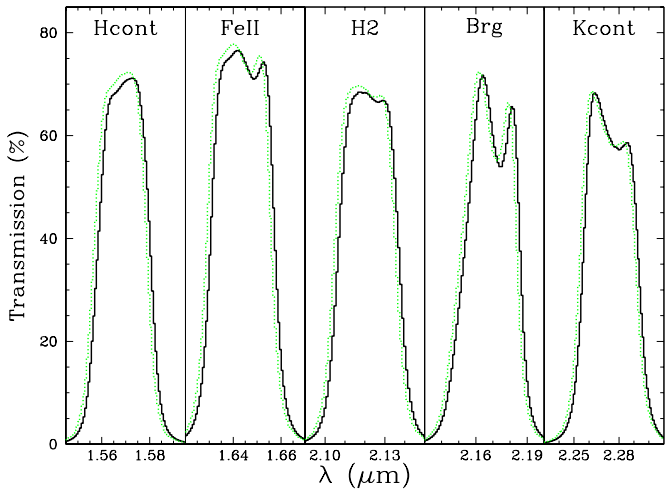}}}}
\caption{
Detailed view of the narrow band filter curves. The solid line
is for an incidence angle of 3$^o$ while the dashed curve refer to 
measurements at 6.5$^o$. The shift in wavelength is compatible with an
effective refractive index of 2 for all filters.
}
\label{fig2}
\end{figure}

The measured transmissions of the  
MKO near-infrared consortium filters 
mounted in NICS are displayed in Fig.~\ref{fig1} which also includes, for
comparison, the curves relative to the standard J filter (see the caption
for details). The corresponding half-power points and average transmissions
are summarized in Table~\ref{tab1}.

Fig.~\ref{fig2}
 is a zoom on the transmission curves of the narrow band 
filters and includes measurements taken at different incidence angles.
The shift of the central wavelength and half power points
can be well represented by the relationship

\begin{equation}
\label{eq1}
 \lambda(\theta)\  \simeq \ \lambda(0^o)\  \cdot \ 
     \sqrt{1-{\sin^2\theta\over n_{\rm eff}^2}}
\end{equation}
where $n_{\rm eff}$ is the effective refraction index. Within the range of
angle spanned by our measurements, all the data 
are compatible with $n_{\rm eff}\!\simeq\!2$, a quite typical value for
IR interference filters (see e.g. Vanzi et al. \cite{vanzi})

\section{Conclusions}

This paper reports measurements of 
the spectral transmission
at 78 K of 4 broad-band filters of the Mauna Kea Observatories (MKO)
near-infrared filter set and 5 additional narrow-band filters.
The measured curves confirm the good quality of most of the filters
and could be useful to all those observatories who are, or
will be using the same filters.
The transmission curves are also available in electronic form on the 
TNG web pages ({\it www.tng.iac.es/IRfilters.html}).
We plan to repeat these measurements 
roughly every year to monitor the aging of the coating layers (if any),
the results will be made available on the same web pages.

\begin{acknowledgements}
We are grateful to Alan Tokunaga for all his work and efforts in organizing
the consortium and taking care of the contacts with
the manufacturing companies.

This paper is
based on observations made
  with the Italian Telescopio Nazionale Galileo (TNG) 
operated on the island of La Palma by the Centro Galileo
 Galilei of the CNAA (Consorzio Nazionale per l'Astronomia e l'Astrofisica) 
at the Spanish Observatorio del
Roque de los Muchachos of the Instituto de Astrofisica de Canarias.
\end{acknowledgements}

\end{document}